\documentclass[aps,prl,twocolumn, 10pt,notitlepage,nofootinbib,showkeys]{revtex4-1}
\usepackage{amstext,amsmath,amssymb,amsfonts,bbm, amsthm}
\usepackage[latin1]{inputenc}
\usepackage{fancyhdr}
\usepackage{graphicx}
\usepackage{hyperref}

\usepackage{epstopdf}
\usepackage{rotating}
\usepackage{cleveref}

\usepackage{float}

\usepackage{cases}
\usepackage{tabls}

\usepackage{subfigure}

\usepackage{amsthm}

\usepackage{tocvsec2}

\usepackage{color}

\def\beq{\begin{equation}}
\def\be{\begin{equation}}
\def\ee{\end{equation}}
\def\bes{\begin{eqnarray}}
\def\ees{\end{eqnarray}}

\def\mean{\textrm{mean}}

\def\ab{{\overline{\alpha}}}

\begin{document}

\title{\large \bf Minimum Relative Entropy Distributions With a Large Mean Are Gaussian}

\author{Matteo Smerlak}\email{msmerlak@perimeterinstitute.ca}
\affiliation{Perimeter Institute for Theoretical Physics, 31 Caroline St.~N., Waterloo ON N2L 2Y5, Canada}

\date{\small\today}

\begin{abstract}
We consider the following frustrated optimization problem: given a prior probability distribution $q$, find the distribution $p$ minimizing the relative entropy with respect to $q$ such that $\textrm{mean}(p)$ is fixed and large. We show that solutions to this problem are asymptotically Gaussian. As an application we derive an $H$-type theorem for evolutionary dynamics: the entropy of the (standardized) distribution of fitness of a population evolving under natural selection is eventually increasing. 
\end{abstract} 
\maketitle

\section{Introduction}

Relative entropy (aka Kullback-Leibler divergence) is the central concept of information theory \cite{Kullback1959}. Given two probability distributions $p$ and $q$, the relative entropy of $p$ with respect to $q$,
\begin{equation}
	D(p\Vert q)\equiv \int p(x)\ln\frac{p(x)}{q(x)}\,dx,
\end{equation}
measures the difference in information content between the (prior) distribution $q$ and the (posterior) distribution $p$. As a consequence of Jensen's inequality, $D(p\Vert q)\geq0$ with equality iff $p=q$. When $q$ is uniform and $x$ is discrete (resp. continuous), $D(p\Vert q)$ reduces to (minus) the Shannon (resp. Gibbs) entropy $S(p)$. 

As first articulated by Jaynes \cite{Jaynes:1957fy}, minimizing $D(p\Vert q)$ with respect to $p$ under constraints is a powerful epistemological principle, leading to robust predictions with minimal input. This inference rule can also be motivated purely axiomatically \cite{Shore:1980ky}. On top of its foundational position in statistical mechanics, the Jaynes minimum relative entropy  principle has been successfully applied to countless practical problems in virtually all fields of science \cite{Buck:1991td}. Relative entropy literally attracts human attention \cite{Itti:2009ur}. 

Here we consider the following version of Jaynes' problem: given a distribution $q$ supported on the real line, find the distribution $p$ such that $D(p\Vert q)$ is minimum under the constraint that 
\begin{equation}
	\mean(p)=\mu
\end{equation}
for some constant $\mu$. We show that, if the solution exists for any $\mu$, then this solution is asymptotically Gaussian as $\mu\to\infty$. Moreover the rate of convergence to the Gaussian is determined by the tail behavior of $q$ in a simple, explicit way. 

Our original motivation for investigating this problem is from evolutionary theory \cite{Smerlak:2015vg}. In this context one is interested in characterizing the evolution of a population's distribution of fitness as a function of time (or generation number). As we shall discuss in the second part of this paper, the asymptotic Gaussianity of mean-constrained minimum relative entropy distributions implies an $H$-type theorem for evolution: provided the population is sufficiently large and diverse, the entropy of (standardized) fitness distributions is eventually increasing under natural selection. Another, more elementary application to driven Brownian motion is also given for illustrative purposes.

\begin{figure*}
\includegraphics[scale=.5]{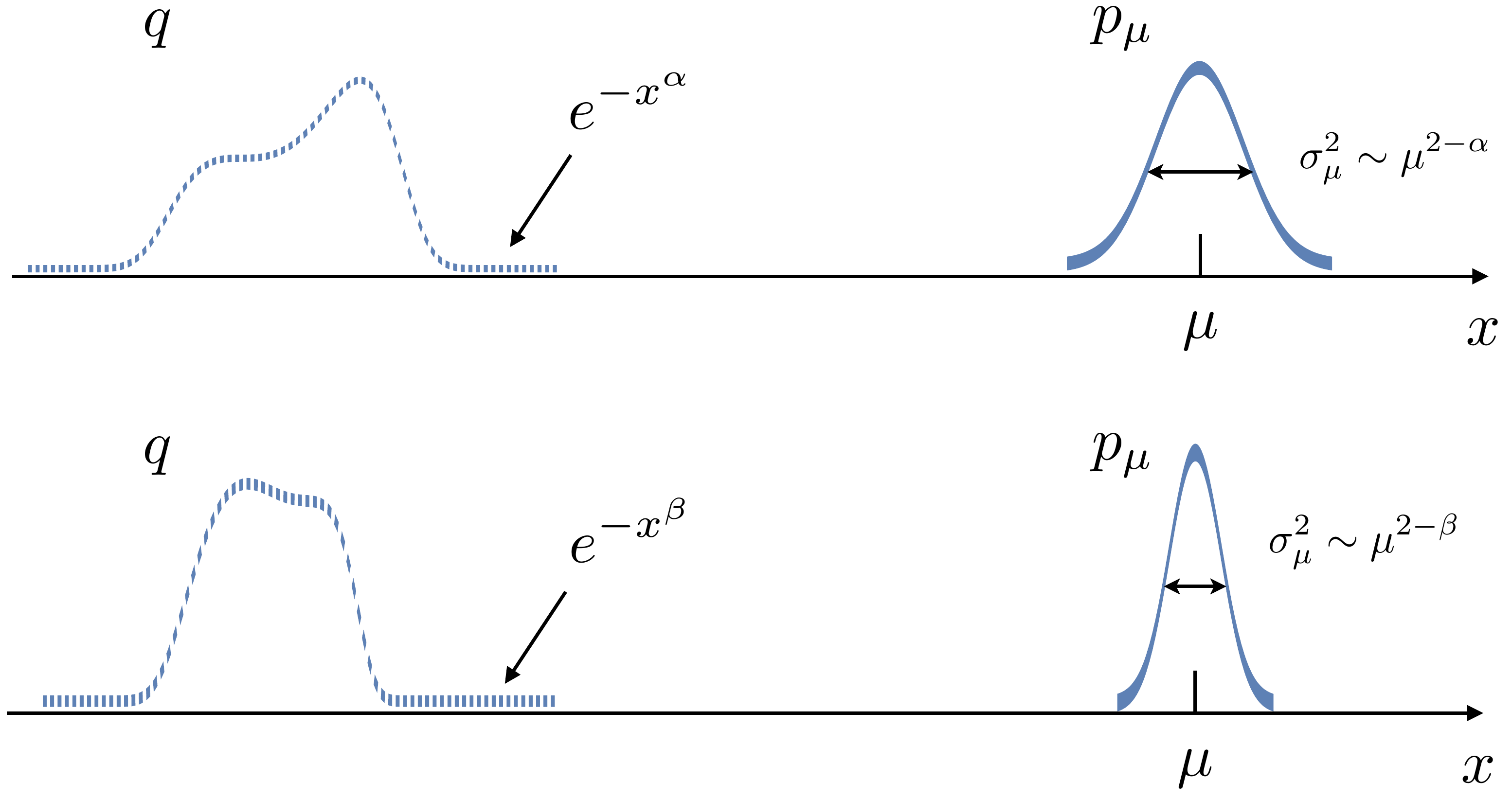}
	\caption{Universality of mean-frustrated minimum relative entropy distributions. The minimizers $p_\mu$ (continuous lines) of $D(\,\cdot\,\Vert q)$ with a given (large) mean $\mu$, for two different priors $q$ (dashed lines). All such minimizers are approximately Gaussian; the only feature distinguishing them is their variance $\sigma_\mu^2$, which is determined by $\mu$ and the tail decay of $q$ according to \eqref{variance}.}
\end{figure*}

\section{Main result}

Given a probability distribution $q$ over the real line, it is well known that the minimizer of $D(p\Vert q)$ under the constraint that the expected value of some function $g(x)$ be fixed to some value $\gamma$ is $p_\gamma(x)=e^{\lambda_{\gamma}g(x)}q(x)/Z_\gamma$, where the Lagrange multiplier $\lambda_{\gamma}$ is determined self-consistently as a function of $\gamma$ (and $q$) and $Z_\gamma$ is a normalizing factor. In particular, taking $g(x)=x$ (\textit{i.e.} fixing the \textit{mean} of $p$) gives the exponentially tilted distribution\footnote{Expression \eqref{solution} is known alternatively as the canonical ensemble (statistical physics), Cram\'er transform (probability theory), natural exponential family (statistics), Esscher transform (actuarial science) of $q$.}
\begin{equation}\label{solution}
	p_\mu(x)=e^{\lambda_{\mu}x}q(x)/\chi_q(\lambda_\mu).
\end{equation}
Here $\chi_q(\lambda)$ is the cumulant-generating function of the prior $q$ and $\mu$ is the fixed value of the mean of $p_\mu$. The multiplier $\lambda_\mu$ is obtained as the implicit solution of
\begin{equation}\label{mean}
	\mu=\chi_q'(\lambda_\mu)/\chi_q(\lambda_\mu)=\psi_q'(\lambda_\mu)
\end{equation}
with $\psi_q(\lambda)\equiv\ln\chi_q(\lambda)$ the cumulant-generating function of $q$. Clearly, the relations above make sense for any $\mu$ only if $q$ decays faster than exponential for $x\to\infty$. To parametrize this decay rate we assume that 
\begin{equation}\label{decay}
	-\ln \int_x^\infty q(x)\,dx\underset{x\to\infty}{\sim}Cx^\alpha
\end{equation}
for some $C>0$ and $\alpha>1$.\footnote{A weaker condition requires that the LHS of \eqref{decay} be regularly varying at infinity with index $\alpha>1$ \cite{Bingham:1989ca}.} Under this condition, the Kasahara Tauberian theorem \cite{Kasahara:1978wda} states that 
\begin{equation}
	\psi_q(\lambda)\underset{\lambda\to\infty}{\sim}D\lambda^{\ab}
\end{equation}
where $\ab=\alpha/(1-\alpha)$ is the exponent conjugate to $\alpha$ and $D\equiv(\alpha C)^{-1/(\alpha-1)}/\ab$. It follows that, in the limit where the mean $\mu$ is large, we have
\begin{equation}\label{tmu}
	\lambda_\mu\underset{\mu\to\infty}{\sim}\alpha C \mu^{\alpha-1}. 
\end{equation}

Let us now show that in this limit $p_\mu$ must be asymptotically Gaussian. Denote $\sigma_\mu$ the standard deviation of $p_\mu$ and let $g_\mu(x)=\sigma_\mu p_\mu(\sigma_\mu x+\mu)$ be the standardized (\textit{viz.} zero mean, unit variance) distribution associated to $p_\mu$. From \eqref{solution}, the $j$-th cumulant of $g_\mu$ is given by
\begin{equation}
	\kappa_\mu^{(j)}=\frac{\psi_q^{(j)}(\lambda_\mu)}{\psi_q''(\lambda_\mu)^{j/2}}.
\end{equation}
Using the Kasahara theorem as above, we have
\begin{equation}\label{cum}
\psi^{(j)}(\lambda)\underset{t\to\infty}{\sim}D(\ab)_j\,\lambda^{\ab-j}
\end{equation}
where $(x)_j=x(x-1)\dots (x-j+1)$ denotes the falling factorial. It follows from \eqref{tmu} and \eqref{cum}	 that the standardized cumulants $\kappa_\mu^{(j)}$ with $j\geq 3$ decrease increasingly fast as $\mu\to\infty$:
\begin{equation}\label{cumulants}
		\kappa_\mu^{(j)} \underset{\mu\to\infty}{\sim}  \frac{[(\alpha-1)C]^{1-j/2}(\ab)_j}{(\ab)_2^{j/2}}\ \mu^{\alpha(1-j/2)}.
\end{equation}
In particular $\kappa_\mu^{(j)}\to 0$ as $\mu\to\infty$ whenever $j\geq 3$, \textit{i.e.} $g_\mu$ converges to the standard Gaussian distribution as announced. Moreover the variance of $p_\mu$ is completely determined by the tail behavior of $q$ (and $\mu$), as
\begin{equation}\label{variance}
	\sigma_\mu^2\underset{\mu\to\infty}{\sim}\frac{\mu^{2-\alpha}}{\alpha(\alpha-1) C}. 
\end{equation}  

A uniform estimate of the rate of convergence can be obtained in terms of the relative entropy $D(g_\mu\Vert\phi)$, \footnote{Bounds on relative entropy are strong: by the Pinsker inequality, the total variation distribution $\delta(p,q)$ between two distribution $p$ and $q$ is bounded as $\delta(p,q)\leq \sqrt{D(p\Vert q)/2}$.} with $\phi(x)\equiv(2\pi)^{-1/2}e^{-x^2/2}$. Denoting $\epsilon_\mu\equiv g_\mu-\phi$ we have
\begin{eqnarray}\label{relent}
D(g_\mu\Vert\phi)&\underset{\mu\to\infty}{\sim}&\int \frac{\epsilon_\mu(x)^2}{\phi(x)}\, dx.
\end{eqnarray}
Now, we can write $\epsilon_\mu$ from the cumulants $\kappa_\mu^{(p)}$ by means of an inverse Laplace transform, yielding
\begin{equation}\label{eps}
		\epsilon_\mu(x)\underset{\mu\to\infty}{\sim}\frac{(\ab)_3\phi(x)(x^3-3x)}{6[(\alpha-1)C]^{1/2}(\ab)_2^{3/2}}\mu^{-\alpha/2}.
\end{equation}
(A more general Edgeworth-type expansion \cite{Feller1979} of $\epsilon_\mu$ on the basis of Hermite polynomial follows similarly.) Plugging \eqref{eps} into \eqref{relent} gives
\begin{equation}
	D(g_\mu\Vert\phi)\underset{\mu\to\infty}{\sim}\frac{(2-\alpha)^2}{6C\alpha(\alpha-1)}\mu^{-\alpha}.
\end{equation}
Thus we see that, the thinner the tail of the prior distribution $q$, the faster the constrained minimizer $p_\mu$ converges to the Gaussian attractor.

We close this section by noting that \eqref{decay} is certainly not the most general condition for $p_\mu$ to be asymptotically Gaussian in the large mean limit. Consider for instance the thin-tailed Gumbel prior $q(x)=\exp(-x-e^{-x})$, a natural distribution in extreme value statistics \cite{deHaan:2007bza}. Then we have $\psi_q(\lambda)=\ln \Gamma(\lambda)\sim_{\lambda\to\infty} \lambda\ln \lambda$, and repeating the computations above shows that $D(g_\mu\Vert\phi)\to 0$ \textit{exponentially} with $\mu$. (This example can be thought of as arising in the limit $\alpha\to\infty$ of the above discussion.)

\section{Representation as transport}

It is interesting to consider the evolution of the shape of the minimizing distribution $p_\mu$ when its constrained mean $\mu$ is varied, or equivalently as the Lagrange multiplier $\lambda$ is varied, as a dynamical system. It is straightfoward to check that the minimizing solution $p_\lambda(x)=e^{\lambda x}q(x)/\chi_q(\lambda)$ satisfies the integro-differential equation
\begin{equation}\label{flow}
	\partial_\lambda p_\lambda(x)=(x-\mu_\lambda)p_\lambda(x).
\end{equation}
Note that, in this dynamical perspective, the prior distribution $q$ is just the initial condition $p_0$ of the flow. Eq. \eqref{flow} can be then used to derive an equation for the standardized distribution $g_\lambda$:
\begin{multline}\label{transport}
	\partial_\lambda g_\lambda(x)-\left(\frac{\ddot{\mu}_\lambda}{2\dot{\mu}_\lambda}\, x+\dot{\mu}_\lambda^{1/2}\right)\partial_x g_\lambda(x)\\=\left(\frac{\ddot{\mu}_\lambda}{2\dot{\mu}_\lambda}+\dot{\mu}_\lambda^{1/2}x\right)g_\lambda(x).
\end{multline}
Here dot means $d/d\lambda$. Thus, the shape of the relative entropy minimizer satisfies a (time-dependent, inhomogeneous) \textit{transport} equation. It can be checked that \eqref{transport} preserves the normalization, mean and variance of $g_\lambda$ as it should. 

The existence of a unique attractor for such a first-order transport equation is somewhat counter-intuitive: we are used to thinking of transport as a non-dissipative process (initial distributions are ``moved around'' without information being destroyed or created). In contrast with this intuition, we have seen that a large domain of initial conditions $g_0$ converge to the standard Gaussian $\phi$ under the transport flow \eqref{transport}. The reason for this behaviour is of course the presence of the ``self-referential'' function $\mu_\lambda$ in this equation: $\mu_\lambda$ is determined by the initial condition $q=p_0$, thereby rendering the problem non-linear. In other words, the function $\mu_\lambda$ captures the shape of the initial distribution in such a way that the time-dependent terms in \eqref{transport} ``erase'' this information over time. 


	\begin{figure}[t!]
	\includegraphics[scale=.9]{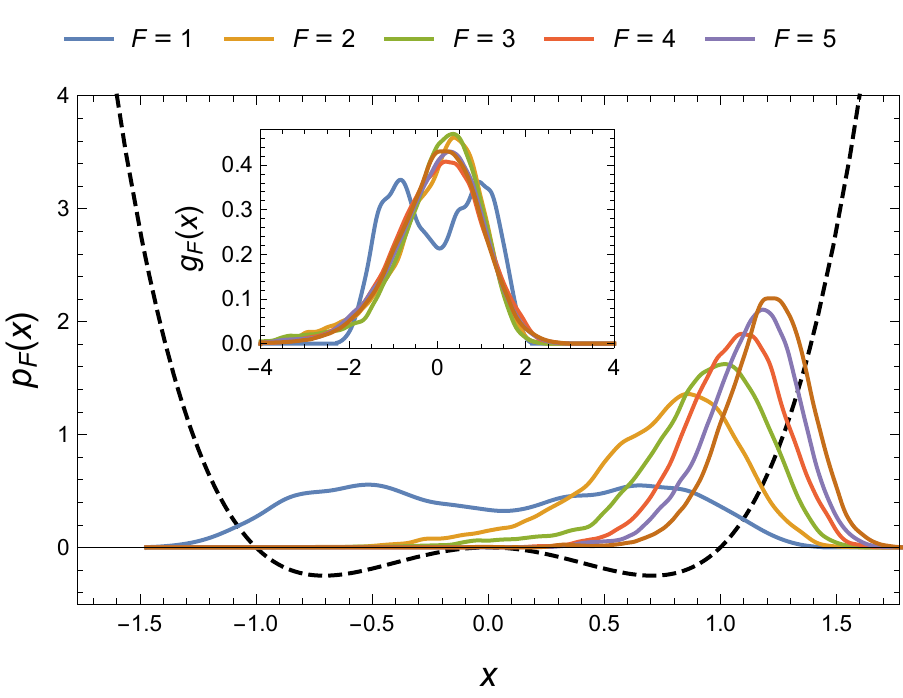}
	\caption{Simulated equilibrium distributions for a Brownian particle in a Mexican hat potential $V_0(x)=-x^2+x^4$ (dashed line) under different applied forces $F$. The inset shows the corresponding standardized distributions $g_F(x)$, which approach the standard Gaussian as $F$ increases. Here $\gamma=1$, $T=1$ and $t\in[0,100]$ with steps $\delta t=10^{-3}$.}
	\label{brownian}
\end{figure}

\section{Applications}  

\subsection{Driven Brownian particle}

As a straightforward application of our limit theorem, consider the overdamped motion of a Brownian particle in one spatial dimension, \textit{viz}.
\begin{equation}
	\frac{dx_t}{dt}=-\gamma V'(x_t)+\xi_t,
\end{equation}
with $x_t$ the position of the particle at time $t$, $V(x)$ a potential, $\gamma$ a friction coefficient, and $\xi_t$ is a Gaussian white noise with $\langle\xi_t\xi_s\rangle=2\gamma T\delta (t-s)$. Assume that $V(x)$ consists of a smooth confining part $V_0(x)$ and of a constant applied force $F$, \textit{i.e.} $V(x)=-Fx+V_0(x)$. Then the equilibrium distribution is 
\begin{equation}
	p_F(x)\propto \exp\left(\frac{Fx-V_0(x)}{T}\right),
\end{equation}
and the results in the previous sections imply that $p_F(x)$ must be Gaussian in the limit of large forces $F$, irrespective of the background potential $V_0$. We illustrate this finding with a Mexican hat potential in Fig. \ref{brownian}.

\subsection{Natural selection}

Let us now consider a different application in the context of evolutionary dynamics \cite{Smerlak:2015vg}. Darwin's principle of the ``survival of the fittest" may be stated as follows: in a population of replicators such that $(i)$ each replicator has a well-defined growth rate (aka ``fitness'') $x$ (\textit{exponential growth}), $(ii)$ not every replicator has the same fitness (\textit{variation}), and $(iii)$ the fitness of descendants is approximately equal to the fitness of parent replicators (\textit{heredity}), then the descendants of the replicators with maximal fitness will eventually take over the entire population, \textit{i.e.} their relative fraction will converge to one. While originally formulated to account for the evolution of biological species,\footnote{Somewhat paradoxically, biological evolution may be the field where natural selection is least strongly established as a dynamical principle. Even condition $(i)$ is hard to verify in real populations \cite{vonKiedrowski:1993dg,Hatton:2015fka}, and it takes experimental engineering to realize exponential replicators in the lab \cite{ColombDelsuc:2015gt}.} this principle is applicable in variety of contexts, from molecules to languages to algorithms to firms. The general relevance of natural selection as an evolutionary force is  referred to as ``Universal Darwinism" \cite{Dawkins:1983fe}. 

A refinement of the principle of the survival of the fittest is Fisher's ``fundamental theorem of natural selection" \cite{Fisher:1930wy}. This celebrated result is the observation that $(i-iii)$ imply that the mean fitness $\mu_t$ in the population grows in time as 
\begin{equation}\label{fisher}
	\frac{d\mu_t}{dt}=\sigma_t^2,
\end{equation}
with $\sigma_t^2$ the fitness variance at time $t$. In particular $\mu_t$ can never decrease under natural selection. Fisher compared this fact with the second law of thermodynamics,\footnote{From \cite{Fisher:1930wy}: ``Professor Eddington has recently remarked that `The law that entropy always increases---the second law of thermodynamics---holds, I think, the supreme position among the laws of nature'. It is not a little instructive that so similar a law should hold the supreme position among the biological sciences."} an analogy which has been hotly debated ever since \cite{Frank:1997ws}. Our result above suggests an alternative heuristic connection between evolutionary dynamics and the second law. Instead of its mean and variance, this new connection involves the entropy of the fitness distribution.

\begin{figure}
	\includegraphics[scale=.3]{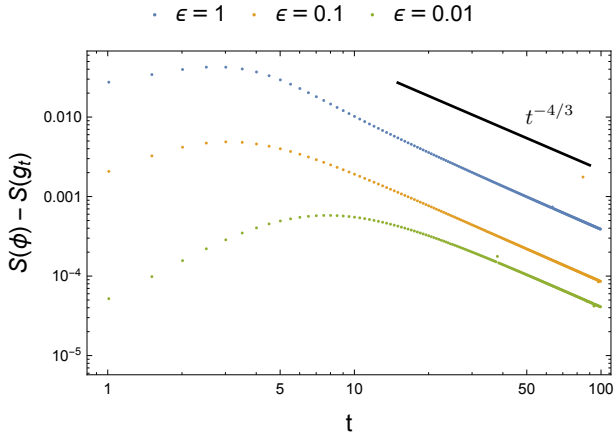}
	\caption{Entropy is eventually increasing but it is not a Lyapunov function for the transport equation \eqref{transport}. Here the initial conditions are $p_0(x)\propto \exp(-x^2/2-\epsilon x^4)$ for three different values of $\epsilon$. Note that, oddly, the closer the initial distribution from the Gaussian (\textit{i.e.} the smaller $\epsilon$), the later the standardized entropy $S(g_t)$ starts increasing towards its limit $S(\phi)$.}
	\label{notLyap}
\end{figure}

Consider indeed a population of replicators such that the density of individuals with growth rate $x$ is $p_0(x)$. Then as a consequence of Darwin's principles $(i-iii)$, we must have after a time $t$
\begin{equation}
	p_t(x)\propto e^{xt}p_0(x),
\end{equation}
\textit{i.e.} the evolved fitness distribution $p_t$ is the minimizer of $D(p_t\Vert p_0)$ with mean $\mu_t$ \cite{Karev:2010gp}. Thus knowing the initial fitness distribution and the mean fitness at all times is equivalent to knowing the entire fitness distribution at all times. Equivalently, $p_t(x)$ is the solution of \eqref{flow} with $\lambda$ as time $t$.   

Now, according to the theorem derived above, provided the population is sufficiently large and diverse so that the support of $p_0$ is effectively unbounded (\textit{i.e.} in a regime of ``positive'' natural selection \cite{Smerlak:2015vg}), the fitness distribution will by force become Gaussian over time. Morover a single ``conserved quantity" (the $\alpha$ tail exponent) completely controls the late-time behavior of the evolving population. Such universality implies that natural selection is a \textit{predictive} hypothesis. That such a system-independent prediction are even possible is sometimes disputed by biologists, who tend to emphasize the ``contingency" of evolutionary changes rather than its universal statistical structure.

To highlight the similarity between the present limit theorem and the $H$ and central limit theorems, it is useful to reformulate our main result in terms of entropy. (We recall that both the central limit theorem and the $H$ theorem are statements about the monotonicity of entropy under the relevant flow---though in the former case this was proved only recently \cite{Artstein:2004wj}). Under the same assumptions as above, we can show that  
\begin{equation}
	S(\phi)-S(g_t)\underset{t\to\infty}{\sim} \frac{(\alpha C)^{1/(\alpha-1)}(2-\alpha)^2}{\alpha-1} t^{-\alpha/(\alpha-1)}. 
\end{equation}

We note that this result is superficially similar to Iwasa's evolutionary $H$ theorem \cite{Iwasa:1988jza}, which identifies a ``free fitness function that always decreases in evolution''. However important differences should be emphasized. First, Iwasa's theorem applies to Markovian models of evolution, and as such it is a result in linear partial differential equations; Eq. \eqref{flow}, by contrast, is a non-linear integro-differential equation without a Markovian interpretation. Second, Iwasa's theorem involves the relative entropy of the probability distribution with respect to a system-dependent final state. Here, on the other hand, the late-time distribution is universal, resulting in a general statistical prediction of Darwin's theory of evolution through natural selection. Third, our result applies to the standardized fitness distribution $g_t$, not to the fitness distribution $p_t$ itself. This is more similar to the entropic central limit theorem \cite{Artstein:2004wj}, which is statement about \textit{rescaled} sums of i.i.d. variables, than to Iwasa's theorem. Fourth, unlike relative entropy for Markov processes, the entropy of $g_t$ is \textit{not} a Lyapunov functional for the flow \eqref{transport}, see Fig. \ref{notLyap}

\section{Conclusion}
Minimum relative entropy distributions with a large mean are asymptotically Gaussian when $\mu\to\infty$. We gave a proof of this result in terms of cumulants, but an alternative, direct-space formulation involving a ``self-referential'' transport equation exists. It would be interesting to understand the dissipative nature of this flow more precisely, for instance by exhibiting a Lyapunov function. 

\medskip

\acknowledgements
I thank C\'edric Villani for a stimulating discussion and for drawing my attention to Ref. \cite{Artstein:2004wj}. Research at the Perimeter Institute is supported in part by the Government of Canada through Industry Canada and by the Province of Ontario through the Ministry of Research and Innovation. 

\bibliography{library}

\end{document}